\documentclass[twoside,twocolumn,9pt]{article}
\usepackage{extsizes}
\usepackage[super,sort&compress,comma]{natbib} 
\usepackage[version=3]{mhchem}
\usepackage[left=1.5cm, right=1.5cm, top=1.785cm, bottom=2.0cm]{geometry}
\usepackage{balance}
\usepackage{mathptmx}
\usepackage{sectsty}
\usepackage{graphicx} 
\usepackage{lastpage}
\usepackage[caption=false]{subfig}
\usepackage[format=plain,justification=justified,singlelinecheck=false,font={stretch=1.125,small,sf},labelfont=bf,labelsep=space]{caption}
\usepackage{float}
\usepackage{fancyhdr}
\usepackage{fnpos}
\usepackage[english]{babel}
\addto{\captionsenglish}{%
  
}
\usepackage{array}
\usepackage{droidsans}
\usepackage{charter}
\usepackage[T1]{fontenc}
\usepackage[usenames,dvipsnames]{xcolor}
\usepackage{setspace}
\usepackage[compact]{titlesec}
\usepackage{hyperref}

\usepackage{epstopdf}

\definecolor{cream}{RGB}{222,217,201}

\begin{document}

\pagestyle{fancy}
\thispagestyle{plain}
\fancypagestyle{plain}{
\renewcommand{\headrulewidth}{0pt}
}

\makeFNbottom
\makeatletter
\renewcommand\LARGE{\@setfontsize\LARGE{15pt}{17}}
\renewcommand\Large{\@setfontsize\Large{12pt}{14}}
\renewcommand\large{\@setfontsize\large{10pt}{12}}
\renewcommand\footnotesize{\@setfontsize\footnotesize{7pt}{10}}
\makeatother

\renewcommand{\thefootnote}{\fnsymbol{footnote}}
\renewcommand\footnoterule{\vspace*{1pt}%
\color{cream}\hrule width 3.5in height 0.4pt \color{black}\vspace*{5pt}} 
\setcounter{secnumdepth}{5}

\makeatletter 
\renewcommand\@biblabel[1]{#1}            
\renewcommand\@makefntext[1]%
{\noindent\makebox[0pt][r]{\@thefnmark\,}#1}
\makeatother 
\renewcommand{\figurename}{\small{Fig.}~}
\sectionfont{\sffamily\Large}
\subsectionfont{\normalsize}
\subsubsectionfont{\bf}
\setstretch{1.125} 
\setlength{\skip\footins}{0.8cm}
\setlength{\footnotesep}{0.25cm}
\setlength{\jot}{10pt}
\titlespacing*{\section}{0pt}{4pt}{4pt}
\titlespacing*{\subsection}{0pt}{15pt}{1pt}

\fancyfoot{}
\fancyfoot[LO,RE]{\vspace{-7.1pt}\includegraphics[height=9pt]{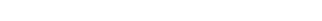}}
\fancyfoot[CO]{\vspace{-7.1pt}\hspace{13.2cm}\includegraphics{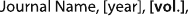}}
\fancyfoot[CE]{\vspace{-7.2pt}\hspace{-14.2cm}\includegraphics{head_foot/RF}}
\fancyfoot[RO]{\footnotesize{\sffamily{1--\pageref{LastPage} ~\textbar  \hspace{2pt}\thepage}}}
\fancyfoot[LE]{\footnotesize{\sffamily{\thepage~\textbar\hspace{3.45cm} 1--\pageref{LastPage}}}}
\fancyhead{}
\renewcommand{\headrulewidth}{0pt} 
\renewcommand{\footrulewidth}{0pt}
\setlength{\arrayrulewidth}{1pt}
\setlength{\columnsep}{6.5mm}
\setlength\bibsep{1pt}

\makeatletter 
\newlength{\figrulesep} 
\setlength{\figrulesep}{0.5\textfloatsep} 

\newcommand{\topfigrule}{\vspace*{-1pt}%
\noindent{\color{cream}\rule[-\figrulesep]{\columnwidth}{1.5pt}} }

\newcommand{\botfigrule}{\vspace*{-2pt}%
\noindent{\color{cream}\rule[\figrulesep]{\columnwidth}{1.5pt}} }

\newcommand{\dblfigrule}{\vspace*{-1pt}%
\noindent{\color{cream}\rule[-\figrulesep]{\textwidth}{1.5pt}} }

\makeatother

\twocolumn[
  \begin{@twocolumnfalse}
{\includegraphics[height=30pt]{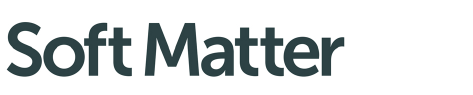}\hfill\raisebox{0pt}[0pt][0pt]{\includegraphics[height=55pt]{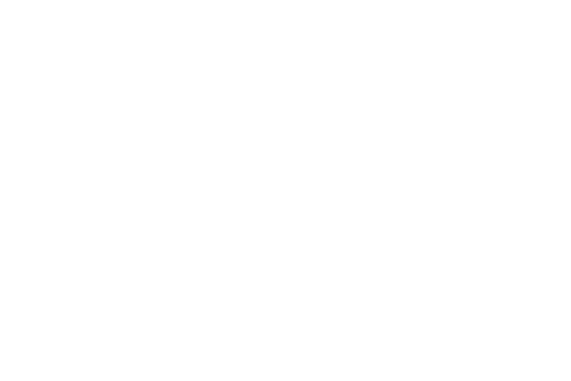}}\\[1ex]
\includegraphics[width=18.5cm]{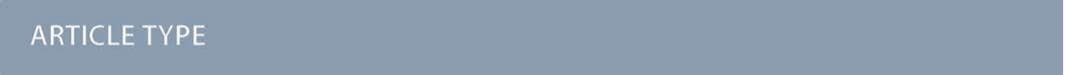}}\par
\vspace{1em}
\sffamily
\begin{tabular}{m{4.5cm} p{13.5cm} }

\includegraphics{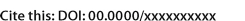} & \noindent\LARGE{\textbf{A molecular dynamics study of surface-directed spinodal decomposition on a chemically patterned amorphous substrate}} \\
\vspace{0.3cm} & \vspace{0.3cm} \\

 & \noindent\large{Syed Shuja Hasan Zaidi,$^{\ast}$ Hema Teherpuria, Santosh Mogurampelly and Prabhat~K.~Jaiswal$^{\ast}$} \\

\includegraphics{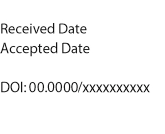} & \noindent\normalsize{We employ a molecular dynamics (MD) study to explore pattern selection in binary fluid mixtures ($AB$) undergoing surface-directed spinodal decomposition on a chemically patterned amorphous substrate. We chose a checkerboard pattern with chemically distinct square patches of a side $M$, with neighboring patches preferring different particle types. We report the efficient transposition of the substrate's pattern as a \emph{registry} to the fluid cross sections in its vicinity when the pattern's periodicity $\lambda/\sigma \simeq 2M$ ($\sigma$ being the fluid particle size) is larger than the mixture's spinodal length scale $\lambda_c/\sigma \simeq 2\pi/\xi_B$ ($\xi_B$ being the bulk correlation length). Our correlation analysis between the surface field and the surface-\emph{registries} in the substrate's normal direction shows that the associated decay length, $L_{\perp}(t)$, increases with decreasing pattern periodicity ($\lambda$). $L_{\perp}(t)$ also exhibits diffusive growth with time $\sim t^{1/3}$, similar to wetting-layer growth for chemically homogeneous walls. Our MD results also show the emergence of composition waves parallel to the substrate, whose wavelength exhibits dynamical scaling with a power-law growth in time $L_{||}(z,t)\sim t^{\alpha}$. $L_{||}(z,t)$ shows dynamical crossovers from a transient \emph{surface-registry} regime to universal \emph{phase-separation} regimes for cross-sections with \emph{registries}. We also give an account of the scaling of \emph{registry's} formation and melting times with patch sizes.} 

\end{tabular}

 \end{@twocolumnfalse} \vspace{0.6cm}

  ]

\renewcommand*\rmdefault{bch}\normalfont\upshape
\rmfamily
\section*{}
\vspace{-1cm}


\footnotetext{\textit{Department of Physics, Indian Institute of Technology Jodhpur, 342037, India. Email:~zaidi.1@iitj.ac.in,~prabhat.jaiswal@iitj.ac.in}}




\section{\label{sec:level1}INTRODUCTION}
Surfaces break the phase separation symmetry, producing directional fluxes, which can govern the evolution of domain morphologies in mixtures during thin-film fabrication or liquid-liquid demixing in micro- or macroscopic confinements. In addition to surfaces, other symmetry-breaking elements, such as impurities and inhomogeneities, manipulate morphologies at various length scales and grow differently from those of pure mixtures.  
Similar types of feature appear for systems with geometrical confinements, such as phase separation in porous media\cite{BG21pre, APD14pre, CSPP20pre, SSC17jcp, SS23tf, SMVBP11jcp} or films,\cite{puri05,KSS10,SSJ06,SSJ06-pre} where symmetry-breaking composition waves from different surfaces dominate bulk phase separation. In this work, we study fluid-fluid demixing on chemically heterogeneous surfaces, where the heterogeneity is decorated as a chessboard pattern with a characteristic length scale or periodicity ($\lambda$). The black and white squares in the pattern are chemically distinct and are selectively wet by a single preferred phase each, from the binary mixture ($AB$). We examine the layer-wise pattern selection by the demixing fluid emerging from an interplay of pattern periodicity and bulk length scales in a semi-infinite geometry.  

A pure binary mixture spontaneously separates during spinodal decomposition (SD) into domains of phases rich in $A$ and $B$ when quenched from a homogeneous state to a state inside the spinodal curve.\cite{VS09,bray02,onuki02} The emergent far-from-equilibrium state is then driven towards its new thermodynamical equilibrium through a series of phase separation stages governed by interfacial tension between the two coexisting phase domains. In the initial stage, low-amplitude high-frequency fluctuations reduce the free energy the most, which develops into the spinodal length scale $\lambda_c$ of the mixture from where the domain coarsening begins. The spinodal length scale could be nanoscopic, microscopic, or mesoscopic for polymer blends ($\sim 100$ nm), colloidal suspensions ($\sim \mu$m ), etc.  During the coarsening stage, the time evolution of the size of the domains strictly depends on factors such as the composition ratio, external fields, global conservation laws, hydrodynamics, and the presence of quenched and annealed disorders.\cite{VS09, tanaka01, BG21pre, APD14pre, CSPP20pre, SSC17jcp, SS23tf, SMVBP11jcp} The domain morphology is characterized by two-point equal-time correlation functions [$C(\Vec{r},t)$] whose decay length yields the characteristic length scale [$L(t)$] proportional to the average domain size. When $L(t)$ is the dominant length scale, the domain patterns become self-similar in time on rescaling by $L$, exhibiting dynamical scaling of the form
\begin{align}
C(\vec{r},t) & =g(r/L),\label{Eq.1}  \\
S(\vec{k},t) & = L^d f(kL),\label{Eq.2}
\end{align}
and the length scale grows in a power law fashion, $L(t)\sim t^\alpha$; where the scaling functions $g(x), f(x)$ in the correlation function $C(\vec{r},t)$ and its Fourier transforms $S(\vec{k},t)$, respectively, and the scaling exponent $\alpha$ are universal. 


	However, phase separation in the presence of surfaces and their unmatched preference for any particle species in the binary mixture $AB$, say $A$, leaves the substrates completely wet by eliminating any contact lines between the two coexisting phases from being in contact with the surface. The phenomenon of wetting on the surface together with spinodal decomposition is termed as surface-directed spinodal decomposition (SDSD) and has been studied extensively in experiments,\cite{MRA95,RLE91,MG03,krausch95, PA91,APF92,BCA93,Michels11,YDL11} analytical,\cite{SK92,troian93prl,troian92,HPW97,RR90,KH91,marko93} and simulations.\cite{SY88,SH97,puri05,binder83,GA92,SK01,SK94,PSS12,SSJ01,SSJ05,SSJ06-pre,bray02,APF21,toxvaerd99prl,HA97pre,PWA94prl,HT00,tanaka01,SAR20,PPS20,PPS20-2,MPJ16} The first successful numerical investigation of SDSD was by Puri-Binder (PB), who studied a coarse-grained model of Cahn-Hilliard-Cook (CHC) supplemented with two crucial boundary conditions for the surface.\cite{SK92} For the completely wet ($CW$) case, the surface becomes the origin of the composition waves known as the SDSD wave, which are composed of alternating wetting and depletion layers of the preferred component. The SDSD waves show damped oscillations near the surface and slowly decay in the bulk while encountering the randomly oriented spinodal waves from all directions. Such SDSD waves are ubiquitous and form a universal feature in experiments.\cite{RLE91,MRA95,YDL11,MG03,PA91,APF92,BCA93} The two featured studies of SDSD are the temporal evolution of the wetting layer, $R_1(t)$, and the domain morphologies in cross sections parallel to the surface. Since a surface has broken the symmetry near the surface, the domain morphology is dominated by the newly introduced length scale $R_1(t)$ perpendicular to the surface. Such anisotropic effects of surfaces lead to slow growth due to nucleation in off-critical mixtures,\cite{puri05,GEF94,SK01} growth arrest in porous medium,\cite{TanakaShimizu17sad,NW21jcp} and also result in elongated domains, oriented parallel to the surface, with enhanced growth. \cite{puri05,tanaka01,SK01}  

Investigation of phase separation in bulk systems and at surfaces is intriguing. Meanwhile, recent focus has shifted towards the phenomenon of phase separation occurring on patterned surfaces. These patterned surfaces exhibit well-defined, geometrically distinct zones, each possessing unique chemical or physical characteristics. The study of phase separation in binary mixtures on patterned surfaces has garnered significant attention in research due to its wide range of applications. Chemically patterned surfaces have been the focal point of numerous studies in various fields, including investigating droplet morphology on chemically patterned substrates and applying this understanding in manufacturing biochips, among other areas.\cite{WYS} Patterned surfaces find applications in various areas, such as open-air small channel devices, controlled-release drug coatings, and water harvesting.\cite{ZLMC,SMC}Other typical applications are elastomeric stamps for stamping and contact printing, lubrication, and paper industries.\cite{AAG94lang,RJG95scien} Practically, chemically patterned surfaces affects the domain morphologies in thin polymer films with the combination of film's thickness, pattern's periodicity and inherent length scale of phase-separating mixtures.\cite{CDS05,KAJB,BN98,CK00} Additionally, patterned surfaces are also valuable in the pharmaceutical and cosmetics industries, notably in microfluidics, where they guide minute quantities of liquids adhering to these chemical microstructures.\cite{CB99,LP20}Interestingly, nature offers numerous examples of patterned surfaces that we can examine. For instance, we can observe the self-cleaning system exhibited by lotus leaves, rice leaves, butterfly wings, and other structures.

	The wetting over chemically patterned substrates for phase-separating binary mixtures has been studied experimentally and numerically to investigate the degree of pattern transpositions in the films as a function of the film's thickness and other essential factors. Karim \emph{et al.}\cite{KDLGRAW98pre} show in their experiments how surface energy modulations created checkerboard patterns and further reported similar qualitative results using simulations to solve CHC equations. In the other three significant works,\cite{ENDRK98prl,NEDK99macro,BWMKS98NATURE} similar groups analyzed the role of pattern periodicity $(\lambda)$, the coarsening length scale $(L(t))$, and the inherent length scale ($\lambda_c$) of the phase-separating mixture. They reported that a fundamental excitation mode dominates when commensurate due to the coupling of $L(t)$ and the surface-deformation modes. Other modes decay or have a very low amplitude. Generally, the studies were performed for thin ($\sim 2000$ nm) and ultrathin films ($\sim 100-300 $ nm), analyzing the dependence of growth modes with film thickness in an attempt to produce films with the desired domain morphology. An analogous study belongs to Chen and Chakrabarti\cite{CC98jcp} on microphase separation in block copolymers (BCPs) with mesoscopic length scales on patterned substrates. They undertook a simple variant of CHC\cite{OB88prl,OS87mplb} to model BCPs (molecular size of $10-100$ nm). The intent was again to study the commensuration between the bulk unconstrained lamellar size $\lambda_c$ and the linear size of the surface inhomogeneities $\lambda$, and the resulting film morphology. 

    	There are a few recent numerical studies that need to be addressed.\cite{WD06jcp,DPZ13macro,CLXQF13macro,SPZA16macro,NPNKWT17sr,XZWZ19pccp,ZL19pccp,OP87prl,OP88pra,PO88pra} Dessi \emph{et al.}\cite{DPZ13macro} and Serra \emph{et al.}\cite{SPZA16macro} conducted a cell dynamics simulation to study SDSD of BCP over chemically patterned substrate with a modified version of CHC.\cite{CC98jcp} They analyzed the emergent structures due to the interplay of self-aggregating BCPs and pattern transposition. Similarly, dissipative particle dynamics were performed by Xiang \emph{et al.}\cite{XZWZ19pccp} to study the structural transitions in BCPs on chemically patterned substrates. Chen \emph{et al.}\cite{CLXQF13macro} used \emph{self-consistent field theory} to investigate the self-assembly of BCP on similar types of substrates. More recently, Das \emph{et al.} and one of us have\cite{PPS20,PPS20-2} numerically solved SDSD on a chemically and morphologically patterned substrate using CHC equations. Their chemical patterns represent a chessboard with black and white square patches, each color preferring one component from a symmetric binary mixture. In the case of morphological patterns, they created posts of different widths and heights on the substrate, creating channels where the fluid could flow. In this work with chemical patterns,\cite{PPS20}, they characterize domain morphologies near the substrate and quantify the pattern dynamics in cross-sections parallel to the surface. Their diffusion-limited model undoubtedly ignored the importance of hydrodynamics at such a mesoscopic length scale.

        All of the studies described above represent our system and the problem statement. Many of these studies characterized the emergent morphologies and discussed suitable ways to solidify them for technological use, predominantly based on thin films. They tried to investigate the role of film thickness in its morphology and explain the possible instabilities arising from hydrodynamics (capillary effects for an ultrathin film with an air/polymer interface) for this reduced dimensionality. Although the recent work by Das \emph{et al.}\cite{PPS20}  presented a detailed pattern dynamics, it lacked hydrodynamics, which accelerates domain growth.\cite{tanaka01} To our knowledge, there are no MD studies in this context, and we can undoubtedly assert the importance of such a study in many ways. Firstly, MD naturally incorporates hydrodynamics, which is crucial in phase-separating critical mixtures, as it accelerates growth in SDSD at various length and time scales.\cite{PA91, APF92,BCA93,ZJPP22pre} Over and above that, there is a need to understand pattern dynamics at length scales comparable to current nanoscopic devices. However, we have found that the results of the current simulation model for bulk phase separation and SDSD on homogeneous walls scale up nicely for polymer blends and colloidal suspensions at microscopic length scales. In this proposition, we extensively performed MD to qualitatively and quantitatively comment on pattern selection in a phase-separating mixture on a chemically patterned substrate. In Sec~\ref{sec:level2}, we describe the methodology used, followed by Sec~\ref{sec:level3}  of results and discussion, and final Sec~\ref{sec:level4} for the summary of our study. 

	\section{\label{sec:level2}METHODOLOGY AND NUMERICAL DETAILS :}
	
	\begin{figure}[!htbp]
		\centering
		\includegraphics[width=\linewidth]{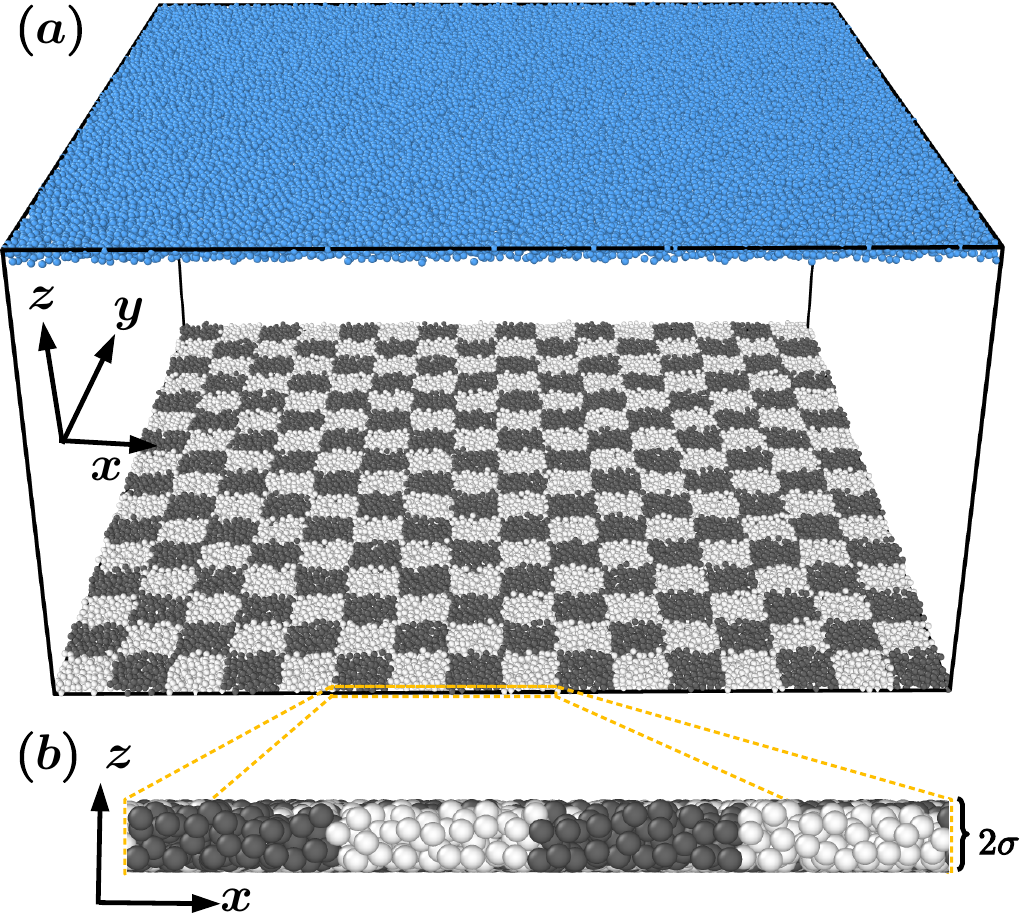}
		\caption{($a$) A $3$D perspective of the simulation box employed in our study of SDSD of a binary mixture ($A+ B$) on a chemically patterned substrate. The simulation box is of size $L_x\times L_y \times L_z = 128\times 128\times 68 \;\sigma^3$ that is periodic in $xy$ direction and confined by amorphous walls in $z$-direction. Fluid particles are not shown here to clearly display the bottom pattern. The substrate pattern at $z=-2$ is a checkerboard type with gray (prefers $A$) and white (prefers $B$) square patches of size $M^2$. An impenetrable wall shown by blue-colored particles is at the top edge $z=64$ of the simulation box that uniformly repels $A$ and $B$. ($b$) A $xz$ cross-section of the chemically patterned wall of thickness $2\sigma$ with $\sigma=1$.   }
		\label{fig:figure0}
	\end{figure}
	
	We performed molecular dynamics simulations of surface-directed spinodal decomposition (SDSD) in a fluid mixture on a chemically patterned substrate using a highly parallel LAMMPS engine.\cite{plimpton95} 
    We considered a binary mixture ($A+B$) in a semi-infinite simulation box periodic in the $xy$ direction and bounded by two amorphous substrates in the $z$ direction. 
    The amorphous substrates were chosen purposely to eliminate the layering effects of a flat wall. We adopted the setup from our previous work of SDSD on the chemically homogeneous amorphous wall and modified it to include the chemical heterogeneity of the substrate.   
	
	Our system is composed of Lennard-Jones particles of mass $m$ interacting with the usual $12\textit{-}6$ LJ potential of the form 
	\begin{equation}\label{eq4:LJ}
	U_{\alpha \beta}(r) = 4\epsilon_{\alpha\beta}\left[ \left(\frac{\sigma_{\alpha \beta}}{r}\right)^{12}- \left(\frac{\sigma_{\alpha \beta}}{r}\right)^6\right],
	\end{equation} where $\epsilon_{\alpha\beta}$ and $\sigma_{\alpha \beta}$ set the interaction strength and \emph{van der Waals} (vdW) radius between different types ($\alpha, \beta\;\; \in \{A, B, S\}$) of particles; $r$ denotes the inter-particle distances. $S$ denotes wall particles that are of the same size as fluid particles. To model phase separation in the fluid and the complete wetting of gray and white patches by $A$ and $B$, we vary $\epsilon_{\alpha\beta}$. We set $\epsilon_{AA}=\epsilon_{BB}=2\epsilon_{AB}=\epsilon$ sufficient to initiate phase separation at the working temperature $T$. The wetting of gray and white patches by $A$ and $B$, respectively, is achieved by setting $\epsilon_{AS_1}=\epsilon_{BS_2}=\epsilon$, where $S_1$ and $S_2$ are particle types belonging to chemically distinct patches. All particles are of equal size and are set to $\sigma$. Also, we truncate and shift the potential in Eq.~\ref{eq4:LJ}, so that both the potential and force approach zero smoothly at the cutoff $R_c = 2.5~\sigma$. \cite{MD17}  Our chosen values for $\epsilon$ and $\sigma$ correspond to a symmetric model belonging to the Ising universality class, which is common in phase separation investigations.\cite{SSS10,SSS12,PSS10,PSS12,PSS12-epl} The patches wetted by a preferred particle type are also neutral to the nonpreferred particles. These neutral interactions, where patches dislike the nonpreferred fluid particles, are modeled by the Weeks-Chandler-Andersen (WCA) potential of the following form:
    \begin{equation}\label{eq3:WCA}
	\phi_{A S_2}(r) = \phi_{BS_1} = 
	\begin{cases}
	4\epsilon\left[ \left(\frac{\sigma}{r}\right)^{12}- \left(\frac{\sigma}{r}\right)^6\right]+\phi_{c},\;\;\; & r<2^{1/6}\sigma\\
	0. & r\ge \sigma
	\end{cases}
	\end{equation}
    
    Moreover, the top substrate is neutral to all particle types. Wall particles do not self-interact, and their positions are not updated in time. That makes them frozen, but they are considered during force calculation for fluid particles.   
	Furthermore, for convenience in simulations, the units of $\epsilon$, $\sigma$, and $m$ are reduced to $1$, yielding a reduced time scale of $t_0=\sqrt{m\sigma^2/\epsilon}=1$. 
	
	We took a simulation box of size $L_x \times L_y\times L_z = 128 \times 128\times 68$, where the fluid is confined in the region of volume $V=128\times 128\times 64$. The top and bottom substrates are each $2$ units thick and are centered at $z=-1$ (chemically patterned) and $65$ (neutral), respectively. The fluid is modeled as incompressible by setting the number density equal to $\rho_N= N/V = 1$, where $N$ is the total number of particles, to avoid coupling to the liquid-gas phase transition. We also keep an equal number of particles of the $A$ and $B$ types, $N_A=N_B=N/2$. A similar number density is kept in the substrate prepared by equilibrating at high temperature, which ensures that the fluid particles do not seep through the walls. 
	
    The simulation protocol includes fluid equilibration at a high temperature for a reasonable time ($\sim 50$ $t_0$), followed by a production run at a lower temperature. During equilibration for $t<0$, the fluid contains only $B$-type particles and does not undergo any phase separation or enrichment on the surface for $T=3 > T_c$ (bulk $T_c \approx 1.423$).\cite{PPS20,PPSD20,puri05,SSS10,SSS12} Instantaneously at $t=0$, the system's thermostat temperature is changed to $T = 1 \approx 0.7T_c$. At the same time, $50\%$ of $B$-type particles are randomly converted to $A$-type. The quenched temperature and critical composition trigger the spinodal decomposition for $t>0$ in the fluid. We then start to record the trajectory of the fluid particles for $t>0$. The temperature of the liquid is controlled via the Nos\'e-Hoover thermostat (NHT),\cite{nosehover,MD17}, and the time integration of the particle's position is performed using the velocity Verlet algorithm.\cite{MD17} NHT has proven to preserve hydrodynamics by conserving momentum and energy locally, producing large-scale fluid flows.\cite{SSS10,SSS12} Due to the nonequilibrium dynamics involved, we did not average quantities over time; instead performed simulations for different initial configurations of the fluid with varying realizations of noise in the thermostat and then performed the statistics on the obtained independent trajectories.   

    Various correlation functions and profiles are constructed on the order parameter $\psi(x,y,z)$, which we define as the local density difference of the particles of $A$ and $B$ type for our system, that is, $\psi(x,y,z)=n_A-n_B$. To compute $\psi(x,y,z)$, we adopt a common coarse-graining procedure used for domain characterizations, where we map the MD simulation box to a lattice structure. We distribute the particles in the MD simulation box into unit cells of sizes $\sigma^3,$ with unique coordinates. Then we compute $\psi(x,y,z)$ in each cell using the number of particles of $A$ and $B$ type present. Furthermore, we remove noise from the raw data using the majority spin rule for $\psi(x,y,z)$, where $\psi$ in a cell is recitified by acknowledging the contributions from neighboring cells. 

\section{\label{sec:level3}Results}

	We begin with a checkerboard of square patches of size $M=16$ having an area of $M_x\times M_y = M^2=16^2$ as shown in Fig.~\ref{fig:figure0}. The checkerboard pattern comprises alternate black and white patches, which prefer $A$- or $B$-type particles, respectively. We have also worked with different patch sizes ($M=8,16,11,32$) to extract the scaling of registry formation and melting times with the size in cross sections transverse to the templated substrate. Moreover, the global composition $\psi_0=(N_A-N_B)/(N_A+N_B)=0.0$ as $N_A=N_B=N/2$ corresponds to a critical mixture, and the area fraction occupied by two types of patches for all patch sizes is $50-50$ of the total surface area of the substrate. An alternate study with incommensurate $\psi_0$ and the template area seems interesting and has been qualitatively reviewed for polymer mixtures.  
	The system is instantaneously quenched at $t=0$ from a homogeneously mixed phase at $T =1.5T_c$ to a spinodal state at $T = 0.7Tc$, where it exhibits spontaneous phase segregation for $t>0$, concomitantly the substrate registers its pattern on the fluid layers in contact.
	
	\subsection{Morphological features of surface registry for a fixed patch size}
	\begin{figure}[!h]
		\centering
		\includegraphics[width=\linewidth]{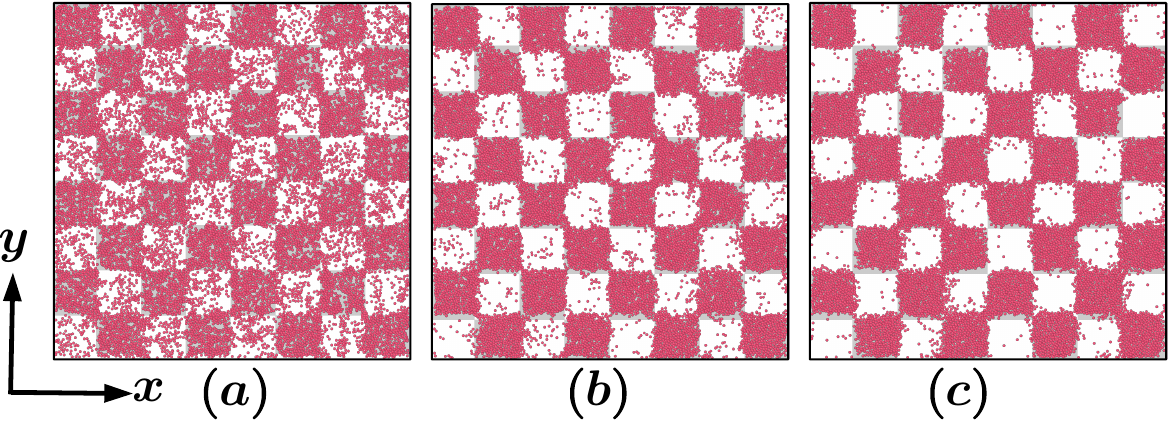}
		\caption{Snapshots of the wetting phenomenon on a chemically patterned substrate at different times. The atomic representations of domains are from the $xy$ cross-sections of width $2$ present next to the patterned substrate and placed upward from $z=0$. The fluid is undergoing SDSD for $T = 0.7T_c$ ( bulk $ T_c \approx 1.423$), which starts from a homogeneously mixed state for $T>T_c$. For clarity, only $A$-type particles (pink) are shown here. As stated in Fig.~\ref{fig:figure0}, the substrate is templated with chemically distinct square patches ($M_x=M_y=16$), and we show here only the projection of gray patches that prefer $A$-type particles. The wetting kinetics of gray patches by $A$ is exhibited by the snapshots belonging to three reduced times of (a) $t=50$, (b) $t=300$, and (c) $t=1000$. }
		\label{fig:figure1}
	\end{figure} 
	
	The template periodicity along the principal axis for $M_x=16$ is approximately five times the spinodal wavelength ($\lambda = 2M \approx 5\lambda_c$), where the inherent length scale of the LJ mixture is $\lambda_c \sim 2\pi/\xi_B \simeq 6.3$ with $\xi_B$ being the bulk correlation length equal to $\sigma$ for LJ symmetric mixtures. For a shorter period along the diagonals, $\lambda = \sqrt{2}M \approx 4\lambda_c$. Along both axes of symmetry, the spinodal length is much shorter than the pattern periodicity, eliminating any commensurability problem between the mixture length scale and template periodicity. Thus, we expect an eminent registry formation in fluid cross sections in contact with the patterned substrate. In Fig.~\ref{fig:figure1}, we show the time evolution of the template's registry in the parallel fluid cross sections of thickness $2$, next to the patterned surface at $z=0$. We only show $A$-type particles that wet the gray patches. The layerwise domain morphology shows that the preferred phase accumulates over gray patches due to wetting, and this wetting overrides pattern selection or macrophase phase separation emerging from the spinodal decomposition (SD). However, this is true only for immediate layers, due to a limited range of the surface potential and large film thickness, where the surface tension overrides the interfacial tension between the coexisting phases. The registry formation in the immediate layers takes some time proportional to the patch area ($M^2$) and has remained intact for the entire simulation time. The registry exhibits domain necking on the template vertices to minimize stresses from interfacial tension. Such domain necking around the corners was also reported in phase-field simulations of the CHC equation under similar circumstances.  
	
	\begin{figure}[!h]
		\centering
		\includegraphics[width=\linewidth]{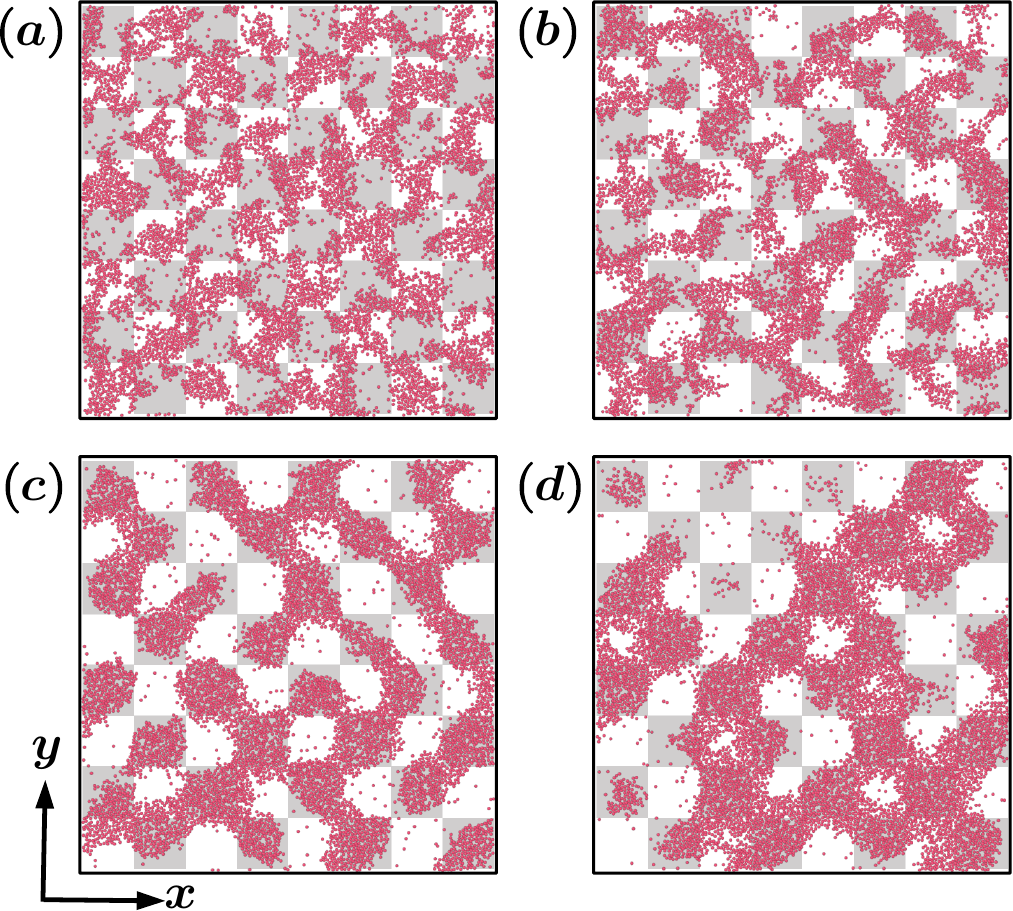}
		\caption{Snapshots analogous to what shown in Fig.~\ref{fig:figure1} but for cross-sections of a unit width starting upward from $z=4$. The time series presented here are (a) $t=500$, (b) $t=1000$, (c) $t=3000$, and (d) $t=10000$. }
		\label{fig:figure2}
	\end{figure}
	
	In Fig.~\ref{fig:figure2}, we present the evolution of the domain morphology from the parallel cross section at a depth of $z=4$. The layers not in contact with the surface are free of surface patterns for some initial time, driven by pattern selection due to SD. The morphology of $t=500$ and $t=1000$ resembles the interconnected morphology of the phase-separating critical mixtures. Later, for $t>1000$, the template registry appears in this layer. The earlier interconnected domain morphology shows a topological makeover to produce the template upwelling from beneath. Notably, the organization is slowed down due to pattern reconfiguring from SD to surface patterns. However, the interfaces are less curved and straighter, forming stripes along the diagonals (to maintain connectivity) and rounded at the edges to minimize interfacial stresses. As time grows, $t>3000$, the layerwise registry enters the transient regime influenced by the macrophase separation, moving towards the bulk morphology of SD. During this regime, we observed that minority domains get stuck in the larger domains of the majority phase, producing impure domains, unlike pure SD domains.

	\begin{figure}[!h]
		\centering
		\includegraphics[width=\linewidth]{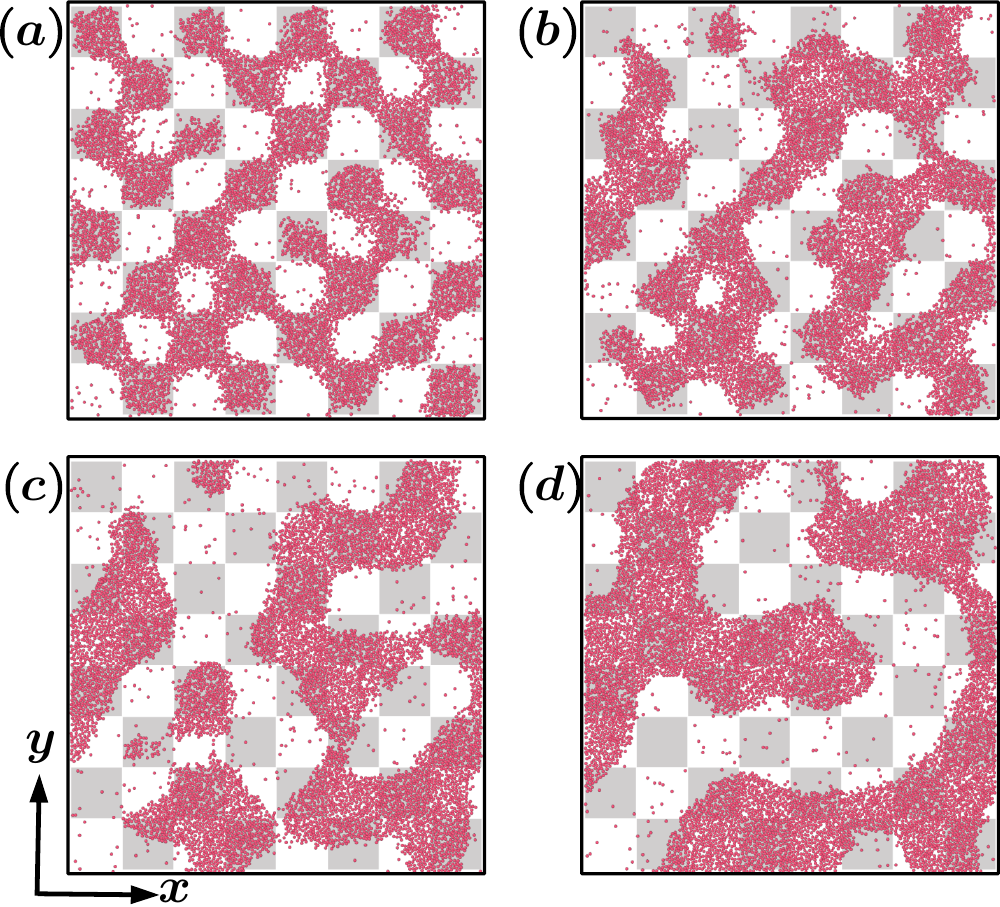}
		\caption{Snapshots with details similar to Fig.~\ref{fig:figure2}, but at different depths from $z=0$ for a time $t=5000$ . The depths denoted by the $z$ values are: ($a$) $2$ , ($b$) $6$, ($c$) $12$, and ($d$) $24$. }
		\label{fig:figure3}
	\end{figure}
	
	Similarly, we examine the registry formed in parallel cross sections as a function of the distance from the patterned surface, shown in Fig.~\ref{fig:figure3}. We present the data for $t=5000$, which lies in the viscous hydrodynamic regime of domain coarsening found in SD and SDSD studies of similar LJ mixtures. By this time, the registry is perfectly transposed to layers at $z=0$ (see Fig.~\ref{fig:figure1}), but the registry is weak in the layers beginning $z=6$, as shown in Fig.~\ref{fig:figure3}(b). The cross sections beyond $z=12$ show no signs of the template registry and exhibit interconnected domain morphology specific to SD. 

    \begin{figure}[!h]
		\centering
		\includegraphics[width=\linewidth]{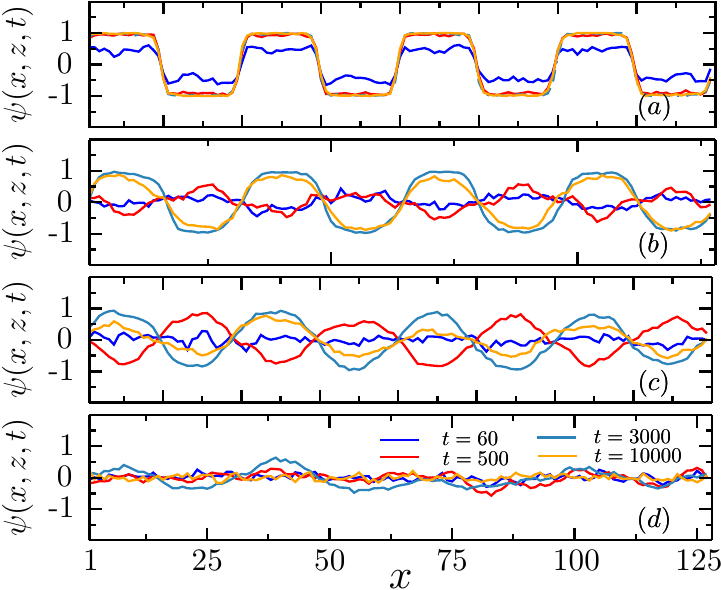}
		\caption{The evolution of the order-parameter profiles $\psi (x,z,t)$ vs. $x$ for times $t$ as specified within the plot. The data are for a fixed $y$ set to $M_y/2=16/2$ but at different $z$ values. Different $z$ values are ($a$) $z=0$, ($b$) $z=2$, ($c$) $z=6$, and ($d$) $z=12$. The temporal profiles are obtained from snapshots presented in Figs.~\ref{fig:figure1}~and~\ref{fig:figure3} after performing the noise removal technique described in the text. 
		}
		\label{fig:figure4}
	\end{figure}
    
    In addition, we perform order parameter profiling in parallel cross sections along the $x$ direction at a fixed $y=L_y/2$, and plot the result in Fig.~\ref{fig:figure4}. Profiling is done using the noise-eliminated coarse-grained simulation box as the methodology explains. Each panel represents the $\psi (\vec{r},t)$ profiles in cross sections centered at different depths $z$, specified in the panels. The registry could be confirmed by the oscillatory profiles shown in Fig.~\ref{fig:figure3}($a$), ($b$), and $(c)$; moreover, the profiles in $c$ also exhibit phase inversions. Phase inversion has been extensively discussed for SDSD over stripped patterned substrates in experiments and continuum simulations of polymer mixtures. Phase inversions emerge from material conservation, yielding depletion layers as a response to the formation of the wetting layer. The depletion layers form a signature of the SDSD wave observed for chemically homogeneous surfaces. The profiles for $z>12$ are flattened at the global composition value $\psi_0=0.0$, highlighting the dominance of isotropic bulk waves of macrophase separation (SD). 

    The registry formation and subsequent phase inversion stem from material conservation in the $z$ direction, as there is no net particle flux. To analyze the correlation between the composition waves and the surface pattern, we define the following quantity: 
    \begin{equation}\label{Eq:g(z)}
        g(z,t)=\frac{1}{L_x\times L_y} \sum_{x,y} (2\Theta(x,y)-1)\psi(x,y,z),
    \end{equation}
    where $\Theta$ is the usual step function
    \begin{equation}\label{Eq:g(z)step}
        \Theta (x,y) =
        \begin{cases}
         1 \;\;\;\;\;\;\;\;\;\;\; x,y \in \text{Gray Patch} \\
         0 \;\;\;\;\;\;\;\;\;\;\; x,y \in \text{White Patch}.
        \end{cases}
    \end{equation}

	\begin{figure}[!htbp]
    \centering
    \includegraphics[width=\linewidth]{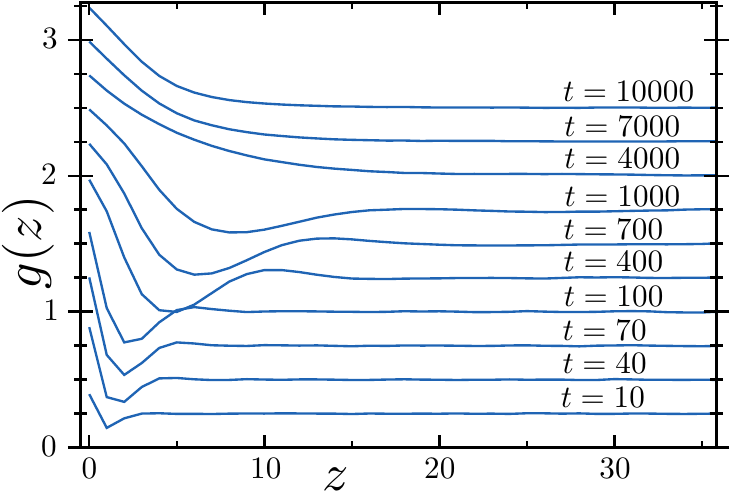}
    \caption{Correlation function $g(z,t)$ for the surface pattern with the patch size $M=16$. After an initial build-up of oscillations till $t\sim 1000$, the oscillations decay again. Each curve has an offset to avoid overlapping.}
    \label{fig:g(z)}
\end{figure}

    We present the resulting composition profiles in Fig.~\ref{fig:g(z)}. The composition profiles show oscillations during early to intermediate time, similar to SDSD waves observed for a chemically homogeneous surface. However, these profiles are more damped with lower amplitudes and fewer oscillations, as each chemically distinct patch becomes the source of oppositely oriented SDSD waves, with neighboring SDSD waves destructively interfering. The composition waves are destroyed in bulk in the usual fashion, arising from the interplay between SDSD and bulk phase separation.  The registry and phase inversion are clear from the zero crossing of these profiles for $t<4000$. Profiles past $t=4000$ are overdamped as macrophase separation melts the registry, beginning with the depletion layers.
    
\begin{figure}[!htbp]
		\centering
		\subfloat{\includegraphics[width=\linewidth]{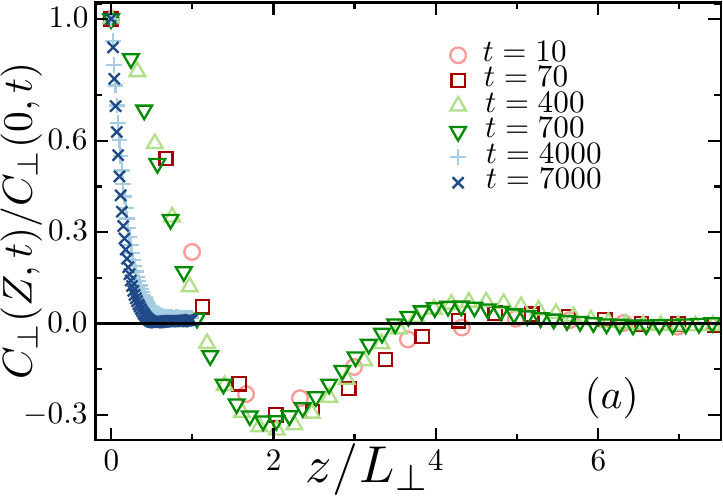}\label{fig:perpC}}
		
		\subfloat{\includegraphics[width=\linewidth]{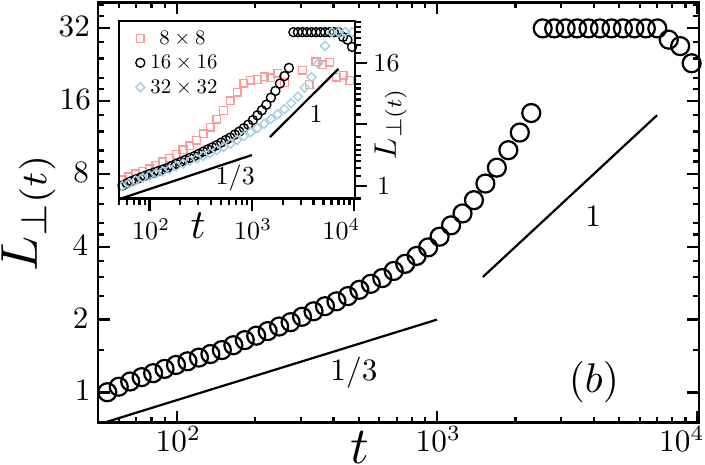}\label{fig:perpL}}
		\caption{($a$) Scaled correlation functions $C_\perp (Z,t)/C_\perp(0,t)$ as a function of scaled distance $z/L_\perp$ normal to the patterned surface. The length scale $L_\perp$ is the decay length and is computed from the first-zero crossing of $C_\perp(Z,t)$. ($b$) Time evolution of the decay length $L_\perp$ on a double-logarithmic plot. $L_\perp (t)$ is compared to the solid lines with known slopes of $1/3$ and $1$ of universal diffusion and viscous hydrodynamics growth regimes, respectively. \emph{Inset} Comparison between $L_\perp(t)$ growth for different patch sizes, $M=8,16$ and $M=32$.}
		\label{fig:perp}
	\end{figure}

To have a scale-independent correlation analysis in an evolving domain morphology, we construct a correlation function from $g(z,t)$ as 
    \begin{equation}
	\begin{split}
	C_{\perp}(Z,t) =  \langle g(z,t)\;g(z+Z,t)\rangle 
	 - \langle g(z,t)\rangle\; \langle g(z+Z,t)\rangle.
	\label{eq:Cperp}
	\end{split}
	\end{equation}     
We plot a normalized and scale-independent version of the above-defined correlation in Fig.~\ref{fig:perpC}. The length scale $L_{\perp}$ is the decay distance of the correlation, which could be assumed to be an approximate wavelength of the oscillations observed for the composition profiles in Fig.~\ref{fig:g(z)}. The rescaled correlations appear to overlap reasonably well for the early to intermediate stages ($t<4000$), highlighting the dynamical scaling of the evolving composition waves $g(z,t)$. The correlations decay rapidly in the late stages, $t>4000$, due to melting of the registry. We further show how the correlation length $L_{\perp}(t)$ scales with time in Fig.~\ref{fig:perpL}. $L_{\perp}$ seems to follow diffusive growth for intermediate stages, $L_{\perp}\sim t^{1/3}$, as observed for the length scale and wetting-layer growth in SD and SDSD for critical mixtures. However, late-stage growth shows a crossover to unknown growth with a growth exponent larger than $\alpha = 1$ of viscous hydrodynamics. Diffusive growth is confirmed from length scale growth for other patch sizes, shown in the inset of Fig.~\ref{fig:perpL}. The crossover appears to be initiated by the melting of registries due to bulk phase separation and does not represent the crossover from diffusive to hydrodynamic, common in pure fluid mixtures and homogeneous surfaces.   

	Next, we characterize domain morphologies in cross sections parallel to the patterned surface by calculating correlation functions at various depths $z$, which we define as:
	\begin{equation}
	\begin{split}
	C(\vec{\rho},z,t) = \frac{1}{L_x \times L_y} \int d\vec{R}\langle \psi (\vec{R},z,t)\psi (\vec{R}+\vec{\rho},z,t)\rangle \\
	- \langle \psi (\vec{R},z,t)\rangle \langle \psi(\vec{R}+\vec{\rho},z,t)\rangle.
	\label{eq:C}
	\end{split}
	\end{equation}
	The angular brackets denote averaging over an ensemble of runs with independent initial conditions and noise realizations. As the system is semi-infinite in the $xy$ plane, we spherically average the correlation function assuming isotropic contributions from the planar coordinates $\vec{\rho}$ to obtain $C(\rho,z,t)$. The dynamical scaling with a general form given in Eq.~\ref{Eq.1}, is written for the layer-wise correlations of Eq.~\ref{eq:C} in an inexact fashion as a function of $z$: 
	\begin{equation}
	C(\rho,z,t) = g_z \left[ \frac{\rho}{L(z,t)} \right]. 
	\label{eq:Cscale} 
	\end{equation}
	$L(z,t)$ is the parallel length scale calculated from the first-zero crossing of $C(\rho,z,t)$ and is proportional to the average domain size. We have also computed structure factor $S(\vec{k}_{\rho},z,t)$, which is simply the Fourier transform of $C(\rho,z,t)$, denoting the in-plane scattering:
	\begin{equation}
	S(\vec{k}_{\rho},z,t) = \int d\rho e^{i\vec{k}_\rho  \cdot \vec{\rho}} \; C(\vec{\rho},z,t).
	\label{S}
	\end{equation}
	
	We again spherically average the structure factor in the $\vec{k}_\rho$ plane to obtain $S(k_\rho,z,t)$. 
	Statistical data are collected from $20$ independent runs.
	
	\begin{figure}[!htb]
		\centering
		\subfloat{\includegraphics[width=\linewidth]{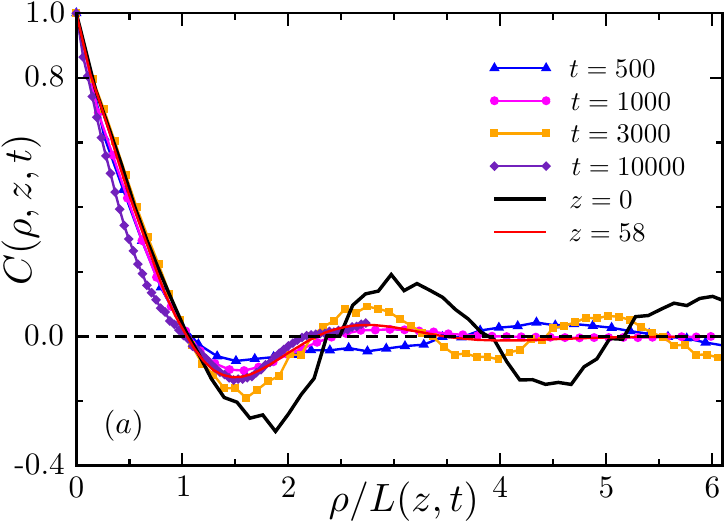}\label{fig:figure8a}}
		
		\subfloat{\includegraphics[width=\linewidth]{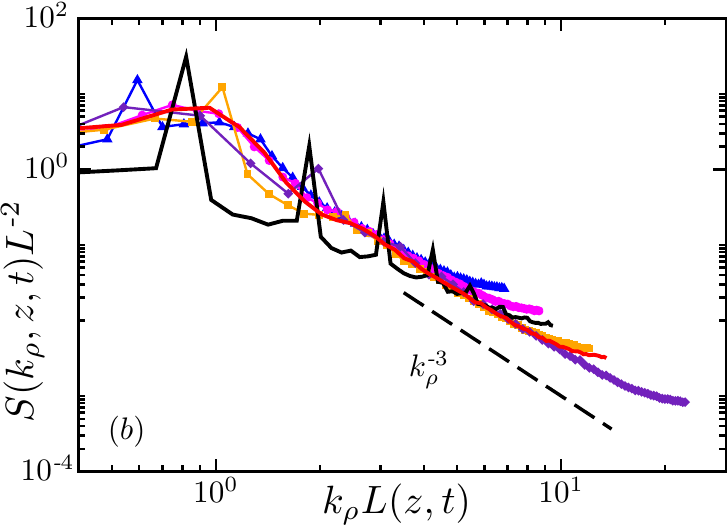}\label{fig:figure8b}}
		\caption{Scaling plot of the layer-wise correlation functions and structure factors for the evolution shown in Fig~\ref{fig:figure2}. ($a$) Plot of  $C(\rho,z,t)$ vs. $t$. The data correspond to four different times at $z=4$. The data for $z=1$ and $z=58$ at $t=3000$ are shown with solid curves as references to the pattern dynamics and the bulk phase separation, respectively. Moreover, we define the characteristic length scale $L(z,t)$ of the fluctuations at $t$ as the first-zero crossing of $C(\rho,z,t)$. ($b$) Log-log plot of $S(k_{\rho},z,t)L(z,t)^{\textit{-2}}$ vs. $k_\rho L(z,t)$ for $z=4$ at different times. The symbols carry similar meanings as above. The \emph{Porod law} is mentioned as a spaced line with the decay form labeled as $k_\rho^{\textit{-}3}$. }
		\label{fig:figure8}
	\end{figure} 
	
	In Fig.~\ref{fig:figure8}, we plotted scaled $C(\rho,z,t)$ and $S(k_\rho,z,t)$ at different times. In Fig.~\ref{fig:figure8a}, the solid red curve represents the data belonging to the bulk cross section at $t=3000$, where the bulk spinodal characterizes the domain morphology. Due to surface-driven fields, the template registry is transposed to fluid layers in the substrate's vicinity, shown as a solid black line for $t=3000$, where the pattern periodicity is characterized by the tail oscillations of $C(\rho,z,t)$. The rest of the data correspond to the domain morphologies analogous to those shown in Fig.~\ref{fig:figure2}, which is a mixture of registry and bulk phase separation. A reasonable overlap is achieved between the bulk layer and the layer at $z=4$ for all times for distances close to the substrate. However, most of the time, the layer-wise domain shows transient morphology, constituting a time series of the formation and the melting of the template registry. Further, in Fig.~\ref{fig:figure8b}, we plot a scaled version of $S(k_\rho,z,t)L^{-2}$ versus $k_\rho L(z,t)$ on a log-log plot. As in Fig.~\ref{fig:figure8a}, there is no reasonable data overlap, and peaks are shifted at different times, possibly due to multiple competing length scales. In the limit of large $k\rho$, $S(k_\rho,z,t) \sim k_\rho^{\textit{-}3}$, the data trend follows the well-known \emph{Porod's law}, which is the consequence of the scattering from the sharp interfaces.

	\subsection{Role of patch size on surface registry}
	
	\begin{figure}[!h]
		\centering
		\includegraphics[width=\linewidth]{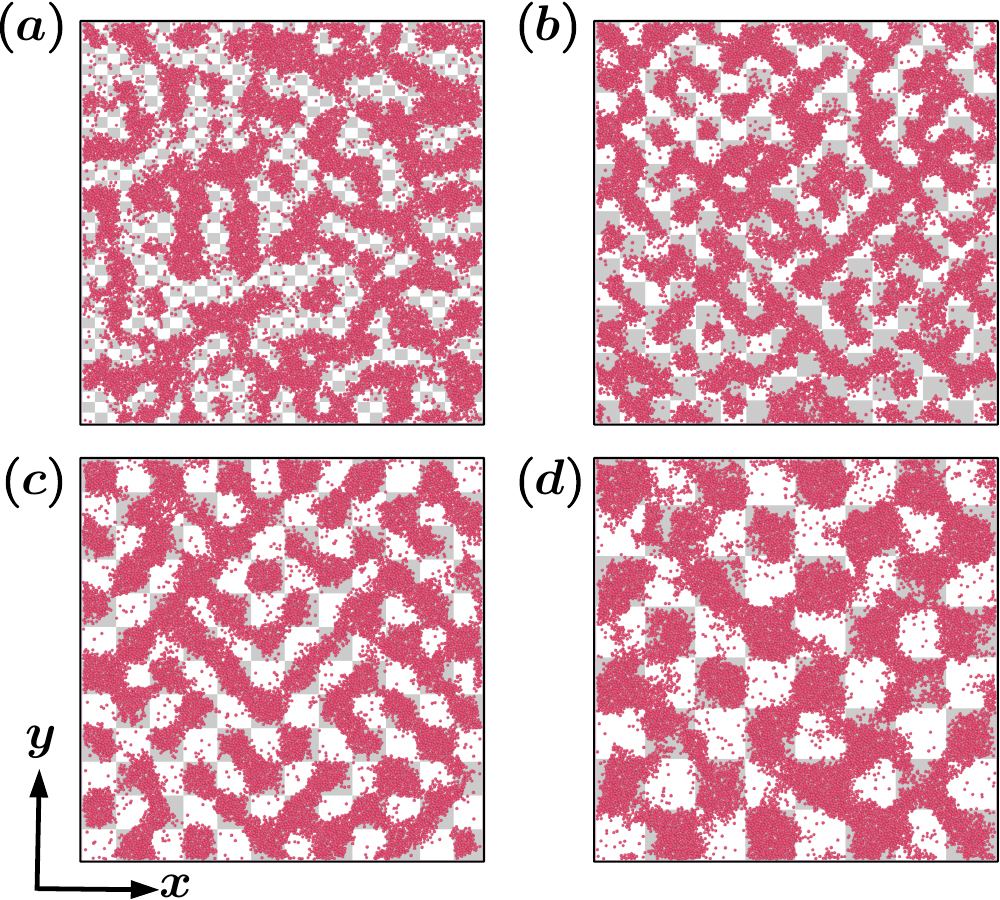}
		\caption{Snapshots analogous to Fig~\ref{fig:figure1} and exhibiting the registry formation at $t=3000$ and $z=2$ for different patch sizes of $(a)$ $M_x \times M_y = 4 \times 4$, $(b)$ $M_x \times M_y = 8 \times 8$,$(c)$ $M_x \times M_y = 10.67 \times 10.67$ and $(d)$ $M_x \times M_y = 16 \times 16$.}
		\label{fig:figure14}
	\end{figure}

	\begin{figure}[!htb]
		\centering
		\subfloat{\includegraphics[width=\linewidth]{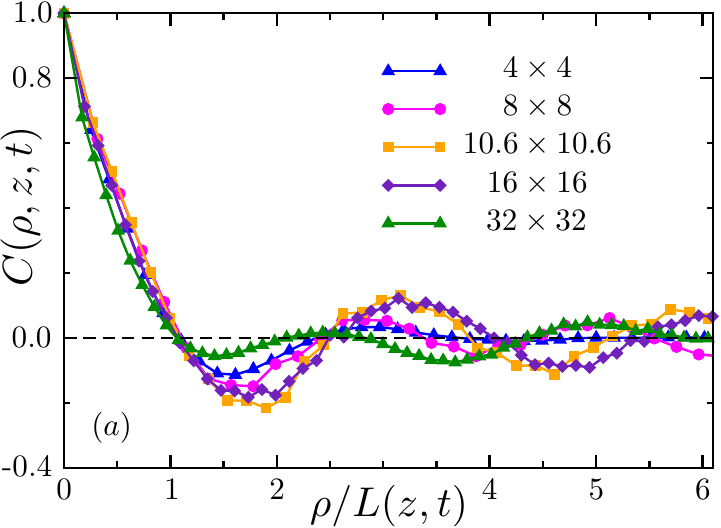}\label{fig:figure15a}}
		
		\subfloat{\includegraphics[width=\linewidth]{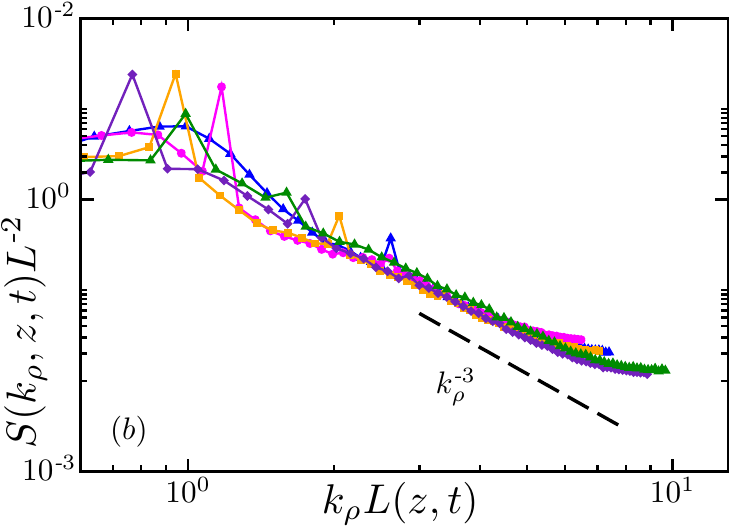}\label{fig:figure15b}}
		\caption{Scaling plot of the layer-wise correlation functions and structure factors analogous to Fig~\ref{fig:figure8}. ($a$) Plot of  $C(\rho,z,t)$ vs. $t$. ($b$) Log-log plot of $S(k_{\rho},z,t)L(z,t)^{\textit{-2}}$ vs. $k_\rho L(z,t)$ for $z=2$ for different $M_x$. }
		\label{fig:figure15}
	\end{figure} 
	
	We explored four more patch sizes in the template $M=4,8,11,32$. The patch size of $M=4$ with $\lambda=8$ is closer to the bulk spinodal length $\lambda_c\simeq 6.3$; therefore, we expect a strong influence of SD on registry formation from the beginning, even for cross sections in contact with the patterned surface. Meanwhile, the wavelength of pattern periodicity for $M_x=32$ is too large, producing outcomes similar to those of a chemically homogeneous surface where neighboring composition waves meet more with waves of similar orientation than oppositely oriented waves. Therefore, we expect a more substantial and deeper transposition of the surface registry in the fluid for $M=32$. We present the domain morphology for different patch sizes in Fig.~\ref{fig:figure14}. The cross-section is at a depth of $z=2$, not in contact with the patterned surface. The surface registry appears only for $M_x\ge 8$ and is absent for $M_x=4$ as expected for $\lambda<\lambda_c$. Therefore, SD influences the surface-driven registry for $M_x=4$, which is still greater than the particle size ($\sigma$). Moreover, we provide more spatial statistics using layer-wise correlations and structure factors for all patch sizes in Fig.~\ref{fig:figure15}. The data are analogous to what is shown in Fig.~\ref{fig:figure8}. In Fig.~\ref{fig:figure15a}, we see bulk behavior for $M_x=4$ with damped oscillation in the tail, whereas other patch sizes of $M_x=8,11,16$ and $32$ show typical surface registry with an oscillating tail of the correlation function. The correlation data for $M_x=32$ show a transient regime in which the registry is still in the early stages. Regardless of the patch size and incommensurability between the spinoidal length and pattern periodicity, the universality of sharp interfaces of the domain morphology is upheld, as seen from the trend in the large wavevectors in the structure factor shown in Fig.~\ref{fig:figure15b} following Porod's law.
		
\subsection{Growth kinetics of surface registry }
	\begin{figure}[!htb]
		\centering
		\subfloat{\includegraphics[width=\linewidth]{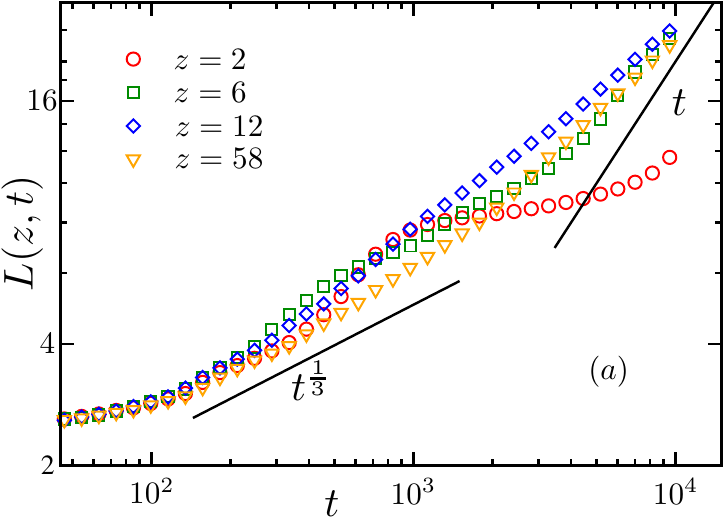}\label{fig:figure11a}}
		
		\subfloat{\includegraphics[width=\linewidth]{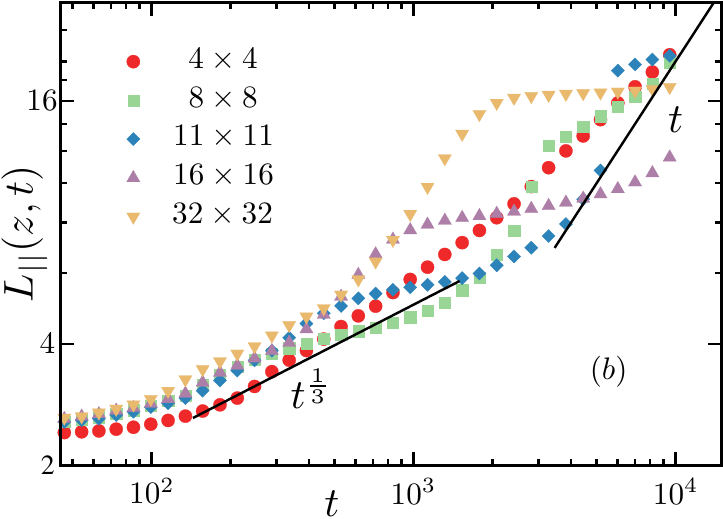}\label{fig:figure11b}}
		\caption{Time-dependence of the characteristic length scale $L(z,t)$. ($a$) Log-log plot of $L(z,t)$ vs. $t$ for different depths $z$ from the substrate as indicated in the plot. The patch size is $M_x \times M_y = 16 \times 16$. ($b$)  Log-log plot of $L(z,t)$ vs. $t$ at $z=4$ and for different patch sizes, as specified therein. The solid lines labeled $t^\alpha$ denote the universal growth regimes of $\alpha=1/3$ (diffusion) and $\alpha=1$ (hydrodynamics).  }
		\label{fig:figure11}
	\end{figure}

	To examine the time evolution of the domain morphology in cross sections parallel to the patterned surface, we extracted the length scale $L_{||}(z,t)$ as the decay length of $C(\rho,z,t)$, and have shown it as a function of time in Fig.~\ref{fig:figure11a} on a double logarithmic scale. The growth kinetics in the surface-field-driven regime show anomalous behavior that saturates at late times after attaining the registry. The cross sections far from the patterned substrate ($z\ge 6$) exhibit diffusive growth ($t^{1/3}$) at intermediate times, marked by a broad crossover to linear viscous hydrodynamic growth ($t^1$) for late stages. Furthermore, in Fig.~\ref{fig:figure11b}, we present $L(z,t)$ for different $M$ at $z=2$ to highlight domain growth under strong and weak registry regimes. SD dominates the growth dynamics for $M=4$ and is not driven solely by surface fields; thus, the length scale closely follows diffusive growth, which finally crosses over to linear growth. 
    The growth remains anomalous in the registry regime for larger patch sizes ($\lambda > \lambda_c$); moreover, the length scales for patch sizes of $M=8,11,$ and $16$ show a melting regime where the registry slowly melts back to bulk domains.   
	
	\subsection{Scaling of formation and melting time of registry with the patch size}
	
	\begin{figure}[!htb]
		\centering
		\subfloat{\includegraphics[width=\linewidth]{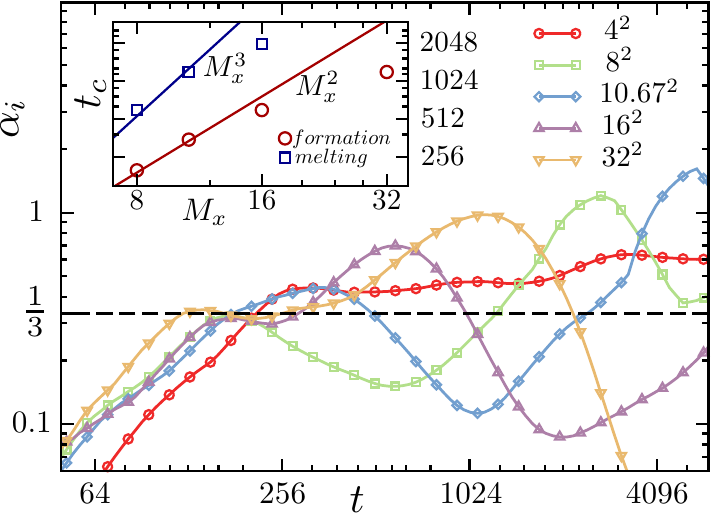}\label{fig:figure12a}}
		
		\subfloat{\includegraphics[width=\linewidth]{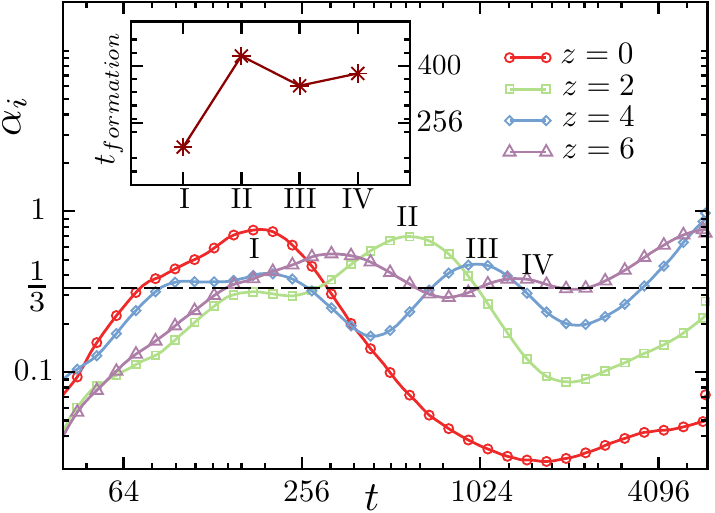}\label{fig:figure12b}}
		\caption{~The log-log plot of the instantaneous exponent $\alpha_i(t)~=~d(\log L(z,t)/d(\log t))$ plotted with respect to time $t$. $(a)$ $\alpha_i(t)$ is extracted from the length scale growth presented in Fig.~\ref{fig:figure11b}. \emph{Inset:} The critical time $t_c$ corresponding to formation and melting of registry. The computation detail of $t_c$ for each $M_x$  is described in the text. The solid lines have slopes of $2$ and $3$ for the scaling of registry formation and melting time, respectively, with $M_x$. 	
			($b$) $\alpha_i$ for $M_x=16$ at different depths $z$. \emph{Inset:} The registry formation time $t_{formation}$ for each peak is computed as the time distance between the current peak and the preceding one.    }
		\label{fig:figure12}
	\end{figure}
	
	As discussed, the transposition of the template registry at any depth depends on the field strength and competing wavelengths of the substrate pattern's periodicity ($\lambda$) and spinodal decomposition ($\lambda_c$). Moreover, we find it intriguing to verify the scaling of registry formation and melting time with patch size $M$. To this end, we extracted approximate times for the formation and melting of registries from the behavior of $L_{||}(z,t)$. In Fig.~\ref{fig:figure12} we plot instantaneous slopes $\alpha_i$ extracted from $L_{||}(z,t)$ of Fig.~\ref{fig:figure11b}. It is obvious from Fig.~\ref{fig:figure11b} that the formation and melting times of the registries correspond to the transition points at which $L(z,t)$ first gradually becomes a plateau, followed by gradual recovery of growth. The plateau is characterized by a sudden and prolonged decrease in $\alpha_i$ as observed in Fig.~\ref{fig:figure12a}. Similarly, recovery due to the melting registries is emphasized by an increase in $\alpha_i$ from its lowest point in Fig.~\ref{fig:figure12a}. 
	
	In the inset of Fig.~\ref{fig:figure12a}, we also show the approximate times of formation and melting of the registries. The registry-formation time scales approximately as $M_x^2$, highlighting its dependence on the surface area governed by the thermodynamics of the near-surface layers. In addition, the outlier corresponding to the formation time for $M_x=32$ is believed to have emerged from hydrodynamics, as it matches the crossover time from the bulk's diffusion to the surface-tension-driven growth regime. In comparison, the melting time of the registry increases as $M_x^3$. The melting time scaling is derived from the fact that $L(z,t)$ grows as $\propto t^{1/3}$, and therefore the registry corresponding to size $M_x$ tends to melt once $L(z,t) \sim M_x$. Thus, the melting time scales as $t_c \propto M_x^3$. Hydrodynamics again produced an outlier in this case, and we were unable to confirm the scaling exponent in this limited study.    
	
	Furthermore, in Fig.~\ref{fig:figure12b}, we plot data analogous to Fig.~\ref{fig:figure12a}, but for a fixed patch size of $M_x=11$. Here, we examine the crossover times for registry formation as a function of distance from the patterned surface. We observed that the formation of registries in layers that do not contact the substrate goes through a multistage process, which increases the formation times (gets doubled), as shown in the inset of Fig.~\ref{fig:figure12b}. The multistage process that causes the delay is the reorganization of the patterns in the required registry from a domain morphology shaped by SD. 
    This is clear from Fig.~\ref{fig:figure2}, where the initial domain morphology in a cross section at $z=4$ develops under SD, showing no signatures of the registry for times up to $t=1000$; however, the registry is present at $t=3000$ without any significant change in the average domain size.

	\section{\label{sec:level4}SUMMARY AND DISCUSSION}
	In this work, we performed molecular dynamics simulations to investigate pattern selection by phase-separating binary fluid mixtures ($AB$) on chemically patterned amorphous substrates. The fluid is modeled as a symmetric Lennard-Jones binary mixture undergoing surface-directed spinodal decomposition (SDSD) in a semi-infinite system that is periodic in the $xy$ direction and bounded by two amorphous substrates in the $z$ direction. The amorphous substrates comprise particles of the same size as the fluid, eliminating layering effects induced by the integrated flat walls in previous studies on SDSD in thin films and semi-infinite geometries. Furthermore, the substrate at $z=0$ exhibits chemically distinct square patches in a checkerboard pattern. Chemically distinct patches prefer $A$ or $B$ from the binary mixture and get wet by them during the SDSD process. We explore the interplay between pattern selection driven by surface fields and spinodal decomposition in directions normal and parallel to the substrate. 
	
	Our MD results show that the substrate pattern is transposed to the fluid as a pattern registry through contacts when the pattern's periodicity ($\lambda$) exceeds the spinoidal length scale of the mixture ($\lambda_c$). The registries appear first in the fluid layers in contact with the substrate, irrespective of the patch sizes for $\lambda > \lambda_c$. The registry is retained in the contact layers during the simulation, dominating the interfacial stresses generated by the bulk phase separation. Also, the highest degree of registry is achieved for such cross sections in contact, which tends to weaken with increasing distance from the patterned surface. 
    
    Our investigation in the $z$ direction shows that the normal composition waves to the surface are weaker in character (less oscillatory and smaller in amplitude) than the SDSD waves obtained for homogeneous walls. The weakening is an outcome of the destructive interference of diametrically opposite composition waves originating from the chemically distinct patches. The resultant composition waves vanish because of randomly oriented composition waves in the bulk. Moreover, we show that the wavelength of these composition waves follows a diffusive growth $L_{\perp}\sim t^{1/3}$ in the intermediate stages of domain coarsening, similar to the growth observed for SDSD waves. The growth becomes robust with decreasing pattern periodicity ($\lambda$). Moreover, when spinodal decomposition starts to melt the registries, $L_{\perp}(t)$ crosses over to an anomalous growth, to which we could not associate any exponent.   
   
   Furthermore, the system reduces interfacial stresses in layerwise registries parallel to the surface by forming a neck between the same-phase domains over the edges of the patches. Also, domains are more rounded at the ends and form rectangular stripes along the diagonal with increasing depth. The square-wave concentration profiles in the contact layers are replaced by sinusoidal oscillations in the non-contact transverse layers. This is verified by the disappearance of kinks in the concentration profiles in non-contact layers along the transverse direction to the surface.   
	
	Our MD results demonstrate that the pattern dynamics in transverse cross sections, when driven by surface fields, show anomalous growth faster than $1/3$ of diffusion. However, the bulk behavior is retained for cross sections not influenced by the surface-driven field. The bulk layer shows diffusive growth ($\sim t^{1/3}$) followed by viscous hydrodynamics ($\sim t^1$). The registries in layers not in contact with the patterned surface display transient behavior for the thickness of the systems considered. The transient registries for some patch sizes form and melt into bulk domains within the simulation time. We show that the formation time scales as the area of patches $t_{formation} \sim M^2$, and melts when the spinodal length scale is of the order of the patch size itself $L_{||}(z,t) \simeq M $. The melting time scales as $t_{melting} \sim M^{1/3}$ during the diffusive regime. We found that hydrodynamics influences the scaling of formation and melting times for larger patch sizes when both time scales are of the order or greater than the diffusive-to-viscous hydrodynamics crossover time observed for SD in pure binary mixtures. Furthermore, transient registries do not exhibit dynamical scaling; however, the \emph{Porod law} dictates that the sharp domain-boundary feature is retained for surface-field-driven and SD growth regimes.






\section*{Acknowledgements}
To be completed \ldots



\balance


\bibliography{test} 
\bibliographystyle{rsc} 

\end{document}